\documentclass[pre,superscriptaddress,showpacs,nofootinbib,
  floatfix,preprint,preprintnumbers]{revtex4-1}

\usepackage[utf8x]{inputenc}
\usepackage{amsfonts}
\usepackage{amsmath}
\usepackage{amssymb}
\usepackage{graphicx}
\usepackage[dvips]{color}
\usepackage{epstopdf}
\usepackage{hyperref}
\usepackage{bm}
\usepackage{upgreek}

\DeclareGraphicsExtensions{.pdf}

\hypersetup{colorlinks=true, citecolor=blue,
  pagecolor=blue,
  urlcolor=black,
  pdfcreator={pdflatex},
}

\begin{document}
\title{Waterlike anomalies in a 2D core-softened potential}

\author{Jos\'e Rafael Bordin}
\email{josebordin@unipampa.edu.br}
\affiliation{Campus Ca\c capava do Sul, Universidade Federal
do Pampa, Av. Pedro Anuncia\c c\~ao, 111, CEP 96570-000, 
Ca\c capava do Sul, RS, Brazil}

\author{Marcia  C. Barbosa}
\email{marcia.barbosa@ufrgs.br}
\affiliation{Instituto de F\'{\i}sica, Universidade Federal
do Rio Grande do Sul\\ Caixa Postal 15051, CEP 91501-970,
Porto Alegre, RS, Brazil}

\begin{abstract}
We investigate the structural, thermodynamic
and dynamic behavior of of a two dimensional core-corona 
system using Langevin Dynamics simulations. 
The particles are modeled using a core-softened potential that exhibits
waterlike anomalies for the 3D case. For quasi-2D systems we have
observed previously a new region of structural anomaly. Now, our results
show that a new region of structural, density and diffusion
anomalies arises for the 2D system. 
Our findings indicates that while the
anomalous region at lower densities observed
is due the competition between the two length scales
in the potential, the traditional mechanism, the higher densities anomalous region is 
related to changes in the particles conformation and a melting region.
\end{abstract}
\maketitle

\section{Introduction}
\label{Intro}

Anomalous materials show characteristics which differ from the observed
in most substances. For instance, it is expected that
liquids  contract upon
cooling at constant pressure and diffuse slower upon 
compression. However, anomalous fluids expand as the temperature is 
decreased and move faster as the pressure grows. 
The most known anomalous system is water, more 
than 70 known anomalies~\cite{URL},
but there are another anomalous fluids.
The  maximum in the diffusion coefficient at constant temperature
was observed not only for  water~\cite{Ne02a} but also for
silicon~\cite{Mo05} and 
silica~\cite{Sa03}. The maximum in the density well know 
in water~\cite{Ke75} is also seen 
in silicon~\cite{Sa03}, silica~\cite{Sh06},
Te~\cite{Th76}, Bi~\cite{Handbook}, 
Si~\cite{Ke83}, $Ge_{15}Te_{85}$~\cite{Ts91},  liquid 
metals~\cite{Cu81}, graphite~\cite{To97} and 
$BeF_2$~\cite{An00}.

Since the seminal work by Jagla~\cite{Ja98, Ja99a, Ja99b} 
core-softened potentials have been widely used in the literature 
to study the 
behavior of anomalous fluids~\cite{Ma05,Sc00a,Xu05, Ol06a, Fomin11, Ya05, Fo08, Lasca10}.
The origin of this
 behavior is associated with the existence of two
characteristic length scales in the potential~\cite{Oliveira07,Barbosa13}.
The competition between the conformation at the first or the second
length scale can be directly related to the anomalies~\cite{Barraz09}.

Core-softened potentials have been also applied to study colloidal systems. 
Experimental works have shown  that the effective interaction between colloids
can be modeled by core-softened potentials~\cite{colloid1, colloid2}.
The origin of the two length scales goes as follows.
The colloids are usually made 
of molecular subunits which form a central packed agglomeration and 
a less dense and more  entropic periferical area.
This core-corona structure can be described  by a 
hard core and a soft corona. Then it becomes natural to  model the
system by a two length scales potential which leads to
the self-assembled
patterns observed in these colloidal 
systems~\cite{Malescio03, Camp03, Forn10, Sin10, Mendonza09, Patta15, Patta17, Patta17b, Scho16, Zhao12, Zhao13}.

Most of works have focused in the self-assembly and distinct patterns
observed in these systems~\cite{Mendonza09, Patta15, Patta17, Patta17b, Scho16,Zhao12,Ciach17, Bordin18a},
with few works focussing on the fluid phase and the dynamics~\cite{Camp03, Zhao13, BoK16a, BoK16b, Bordin16a}.
Therefore, a natural question that arises is how the 
fluid phase of the 2D core-corona system 
behaves for different pressures and temperatures and 
particularly when exposed to a solvent.

In order to address this
question in this paper the interparticle  
colloid-colloid interaction  has a repulsive core with a smooth
shoulder. For a molecular system in 3D this potential 
both in the bulk~\cite{Ol06a, Ol06b} and when confined in
quasi-2D systems~\cite{Krott13, Krott13b, KoB15}
shows waterlike anomalies. Interesting,
in the quasi-2D case a new region of structural anomaly
was observed~\cite{BoK15a}.
Now, we show that 
for 2D systems this
same potential presents a second region of anomalies in the
pressure versus temperature phase diagram.
A mechanism for the appearance of this second anomalous 
region is proposed.

Our paper is organized as follows. In the Section~\ref{Model}  the  model 
and the details about the simulation method are presented. In the 
Section~\ref{Results} results are discussed. The conclusions are
shown in Section~\ref{conclusions}.

\section{The Model and the Simulation details}
\label{Model}

\begin{figure}[ht]
\begin{center}
\includegraphics[width=8cm]{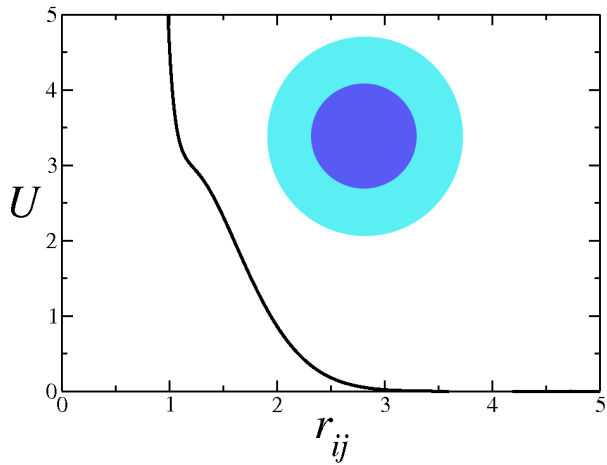}
\end{center}
\caption{Core-softened interaction potential $U$ between two core-corona 
particles. Inset: schematic depiction of the particles, with
the core (first length scale at $r_{ij}\equiv r_1\approx 1.2 \sigma$) and 
the soft corona (second length scale at $r_{ij}\equiv r_1\approx 2.0 \sigma$).}
\label{fig1}
\end{figure}

For simplicity, all the physical quantities are computed 
and displayed in the standard Lennard Jones 
(LJ) reduced units~\cite{AllenTild}.
The system consists of $N = 2000$ disks with 
diameter $\sigma$ and mass $m$ with a potential
interaction composed of a short-range attractive Lennard Jones 
potential and a Gaussian therm 
centered in $r_0$, with depth $u_0$ and width $c_0$, 
 $$
 U(r_{ij}) = 4\epsilon\left[ \left(\frac{\sigma}{r_{ij}}\right)^{12} -
 \left(\frac{\sigma}{r_{ij}}\right)^6 \right] + 
 $$
 \begin{equation}
 u_0 {\rm{exp}}\left[-\frac{1}{c_0^2}\left(\frac{r_{ij}-r_0}{\sigma}\right)^2\right]\;,
 \label{AlanEq}
 \end{equation}
\noindent where $r_{ij} = |\vec r_i - \vec r_j|$ is the distance 
between two disks $i$ and $j$. 
This potential can be parametrized to have a
ramp-like shape, and 
was extensively applied to study systems with 
water-like anomalies~\cite{Ol06a,Ol06b}.
The parameters used in this work are $u_0 = 5.0$, $c = 1.0$ 
and $r_0/\sigma = 0.7$.
The interaction potential, showed in figure~\ref{fig1}, has two 
length scales.
The first scale is at  $r_{ij}\equiv r_1\approx 1.2 \sigma$, where the 
 force has a local minimum, and the other scale at  
$r_{ij}\equiv r_2 \approx 2 \sigma$, where
the fraction of imaginary modes of the instantaneous normal modes spectra has a local minimum~\cite{Charu10}. The 
cutoff radius for the interaction is $r_c = 3.5$.
The two length scales in the potential allow for representing 
hard core-soft shell colloids~\cite{Patta15,Ciach17}.

In this work we use 
the Langevin thermostat~\cite{AllenTild} to mimic the solvent effects.
Hydrodynamics interactions were neglected.
Since the system is in equilibrium we do not expect 
that this will change the long-time
behavior.
The temperature was simulated in the interval 
between $T = 0.01$ and $T = 0.40$.
The number density is defined as $\rho = N/A$, where $A= L^2$ 
is the area and $L$ the
size of the simulation box in the $x$- and $y$-directions.
$\rho$ was varied from $\rho = 0.05$ up to $\rho = 0.60$, and 
the size of the simulation box was obtained via $L = (N/\rho)^{1/2}$.
For clarity, in the $p\times T$ phase diagram 
the higher isochore shown is $\rho = 0.525$ since no anomalous
behavior was observed above this density.

The time step used in the simulations was $\delta t = 0.001$,
and periodic boundary conditions were applied in the two directions.
We performed $3\times10^7$ steps to equilibrate the system. 
These steps are then followed by $5\times10^7$ steps for the results 
production stage. 
To ensure that the system was equilibrated, the pressure, kinetic 
and potential energy as function of time was analyzed.
Snapshots of the system was also used to verify the equilibration.
Also, two distinct initial configurations were used for 
each point: a random fluid-like
configuration and a solid-like in a square lattice.
The results showed independent from the initial configuration.

To study the dynamic anomaly the relation between 
the mean square displacement (MSD) with time was computed,  namely
\begin{equation}
\label{r2}
\langle [\vec r(t) - \vec r(t_0)]^2 \rangle =\langle 
\Delta \vec r(t)^2 \rangle\;,
\end{equation}
where $\vec r(t_0) = (x(t_0)^2 + y(t_0)^2) $ 
and  $\vec r(t) = (x(t)^2 + y(t)^2)$
denote the coordinate of the particle
at a time $t_0$ and at a later time $t$, respectively. The MSD is
related to the 
diffusion coefficient $D$ by 
\begin{equation}
 D = \lim_{t \rightarrow \infty} \frac{\langle \Delta 
\vec r(t)^2 \rangle}{4t}\;.
\end{equation}

The structure of the fluid was analyzed using the radial distribution 
function (RDF) $g(r_{ij})$, and the pressure was evaluated with the 
virial expansion. 
Directly related to $g(r_{ij})$, we characterize the structural 
anomaly using the 
translational order parameter $\uptau$, defined as~\cite{Er01}
\begin{equation}
\label{order_parameter}
\uptau \equiv \int^{\xi_c}_0  \mid g(\xi)-1  \mid d\xi,
\end{equation}
\noindent where $\xi = r\rho^{1/2}$ is the interparticle 
distance $r$ divided by the mean separation between pairs of particles  
$\rho^{1/2}$. $\xi_c$ is a cutoff distance, defined as $\xi_c = L\rho^{1/2}/2$.
For an ideal gas (completely uncorrelated fluid), $g(\xi) = 1$ and $\uptau$ vanishes.
For crystal or ordered fluids a translational long order ($g(\xi) \neq 1$) persists
over long distances, increasing the value of $\uptau$.

In order to check if the system shows density anomaly  the 
temperature of maximum density (TMD) was computed 
for different ischores as follows. Using 
thermodynamical relations, the
TMD was characterized by the minimum in the pressure versus
temperature diagram along isochores,
 \begin{equation}
  \left(\frac{\partial p}{\partial T}\right)_{\rho} = 0\;.
  \label{TMD}
 \end{equation}

The separation between the fluid and amorphous solid phases was 
defined by the analysis
of the total energy, RDF, MSD and system snapshots. When the particles
have a well defined structure
and have a very low or zero mobility the phase
was defined as solid. When the system has nonzero mobility, it
was considered to be in the 
fluid phase. These separations were confirmed by the evaluation of the 
heat capacity~\cite{AllenTild}.
The results were supported by 
larger  simulations, using using $N = 5000$ disks and 
$5\times10^9$ steps.

\section{Results and Discussion}
\label{Results}
\begin{figure}[ht]
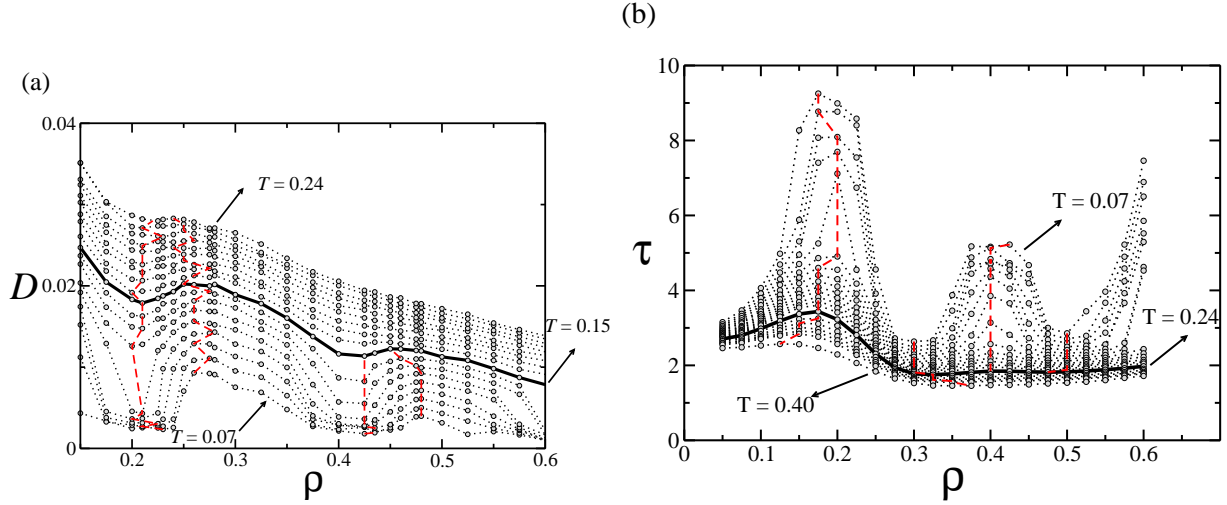

\begin{center}
\includegraphics[width=8cm]{fig2a.eps}
\includegraphics[width=8cm]{fig2b.eps}
\end{center}
\caption{(a) Diffusion constant $D$ and (b) translational order parameter $\uptau$ as function of the system density. In both figures 
maxima and minima that characterize the anomalies are represented by a dashed red line. For the diffusion anomaly, the 
anomalous region at lower densities ranges from the isotherm $T = 0.07$ to $T = 0.24$, while the second anomalous
region goes from $T = 0.07$ to $T = 0.15$. In the case of the structural anomaly, the first anomalous regions goes
from the isotherm $T = 0.07$ to $T = 0.40$, and the anomalous region at higher densities 
is located between the temperatures $T = 0.07$ and $T = 0.24$.
The errors bars in $D$ and $\uptau$ are smaller than the data point.}
\label{fig2}
\end{figure}
For most fluids, the diffusion constant $D$ decreases with 
the density $\rho$. The reason 
for this behavior is that the particles
become more structured as the density increases.
Then the translational order parameter
 $\uptau$,  defined by the Eq.~\ref{order_parameter} grows with
$\rho$ as follows. At low densities, $g(r)\approx 1$ and 
then $\uptau \approx 0$. As 
the density increases, $g(r)\neq 1$ for
many values of $r$ and then  $\uptau$
grows. Anomalous fluids show the opposite behavior. For
these materials in a certain range
of temperature and pressures, the anomalous region, the diffusion coefficient
increases with density, and $\uptau$ decreases with $\rho$.
The figure~\ref{fig2}(a) shows the dependence of the
diffusion coefficient,
$D$, with the density, $\rho$. As the 
density is increased from the 
gas phase, the diffusion coefficient decreases, reaches the
first minimum in the density and  increases reaching a
the  first maximum which
characterizes
the first anomalous region from isotherm $T = 0.07$ to $T = 0.24$. 
Then, as the density
is increased even further, for isotherms between 
$T = 0.07$ and $T = 0.15$, 
a second minimum and a second maximum are observed. 

The translational order parameter versus
density shown in the 
figure~\ref{fig2}(b)also a shows the existence of
 two anomalous regions. The first is located between the isotherms 
$T = 0.07$ and $T = 0.40$ and lower values of density, while 
the second occurs at higher densities 
and from the isotherm $T = 0.07$ to $T = 0.24$.

\begin{figure}[ht]
\begin{center}
\includegraphics[width=12cm]{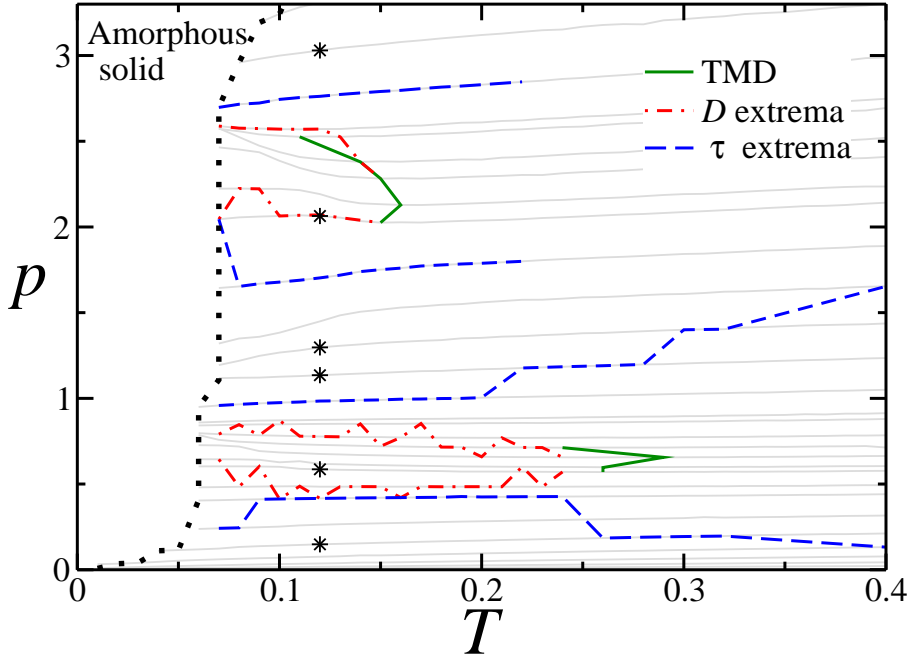}
\end{center}
\caption{$p \times T$ phase diagram of the colloidal system.
The gray lines are the isochores. The dashed blue line delimits the 
structural anomaly
regions, with the maximum and minimum values of $\uptau$. The dotted-dashed red 
line delimits the diffusion anomaly regions, 
with the minimum and maximum values of $D$. The green line
defines the density anomaly region
and corresponds to the temperature of maximum density (TMD) line. The 
black stars are located over the isotherm $T = $ 0.12
and correspond to the densities $\rho = 0.15$, $\rho = 0.225$, 
$\rho = 0.325$, $\rho = 0.35$, $\rho = 0.425$ and $\rho = 0.525$.
The dotted black line delimits the fluid and amorphous solid region.
The errors obtained for the mean value of $p$ and $T$
were smaller than $10^{-4}$ for all cases and the errors bars were omitted
for simplicity.}
\label{fig3}
\end{figure}

In the figure~\ref{fig3} the pressure versus
temperature phase diagram is illustrated for the system. 
The isochores are the gray lines. The temperature
of low density line, related to the 
density anomaly, is shown in green. The TMD
anomaly is also present in both anomalous region.

For the 3D molecular liquid
the TMD region in
the pressure versu temperature 
phase diagram is located inside the diffusion maxima and minima regions
which are
inside the maxima and minima of the translational order 
parameter~\cite{Ol06a, Ol06b} regions. This sequence is the same that 
observed in water, the so-called waterlike hierarchy.
Here, unlike the 3D 
molecular system,
the 2D Brownian system the hierarchy in the anomalies 
is distinct from the water hierarchy.
This change in the hierarchy was already observed in others works, and
it is attributed to the  changes in the competition between
the scales~\cite{Fomin11,Bordin16a, BoK17a}, to the
formation of an ordering 
structure~\cite{Girardi07} or to the dimensional change from 3D to 2D~\cite{Duda14}.

In our case, the change in the hierarchy is 
due to the  presence of solid-like (or
pinning-like)  structures and to the change in the dimensionality
as it is shown next.
In addition to the hierarchy, another question is also 
relevant: why there are
two anomalous regions in this system?

In order to understand the mechanism which generates the 
two anomalous regions, the behavior
of the system at the  isotherm $T = 0.12$,
show as stars in the phase diagram, figure~\ref{fig3}, is analyzed. In the 
case of the 3D system the mechanism which
explains the existence of the waterlike anomalies is  the
competition between the two length scales~\cite{Oliveira07}.
This is observed in the
radial distribution function of molecular systems as follows.

\begin{figure}[ht]
\begin{center}
\includegraphics[width=8cm]{fig4a.eps}
\includegraphics[width=8cm]{fig4b.eps}
\includegraphics[width=8cm]{fig4c.eps}
\includegraphics[width=6cm]{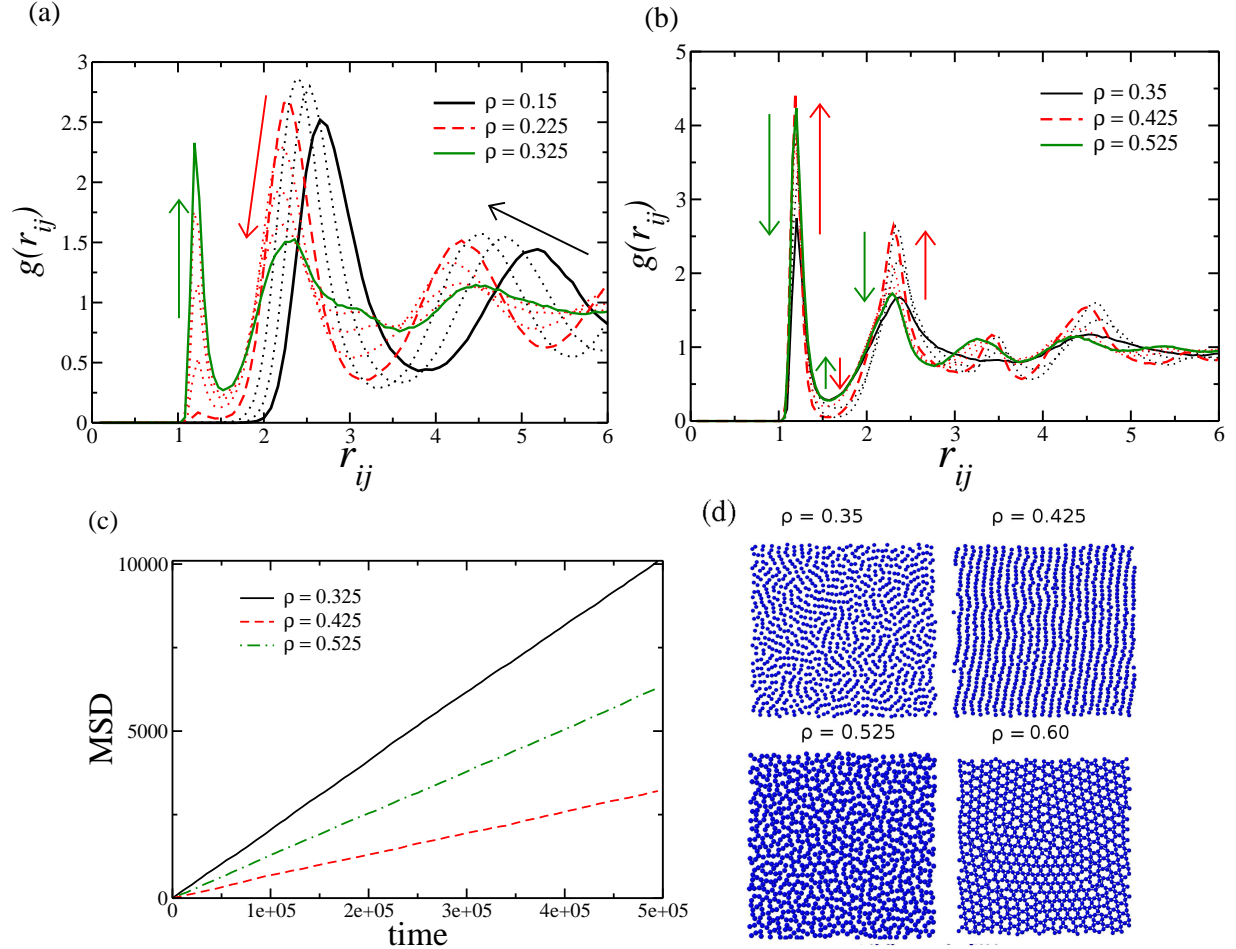}
\end{center}
\caption{Analysis of the isotherm $T =$  0.12 for the colloidal system. 
(a) Radial distribution function (RDF) $g(r_{ij})$ for densities inside the 
first anomalous region indicates that these anomalies are originated by the competition between the two length scales. The 
arrows shows how the peaks in the $g(r_{ij})$ moves. The black arrow shows the grow of the second peak for densities 
below  $\rho = 0.225$, the red arrow the decrease in the second peak and the green arrow the increase in the first
peak for densities between $\rho = 0.225$ and $\rho = 0.325$. (b) Radial distribution function (RDF) $g(r_{ij})$ for 
densities inside the second anomalous region indicates that there is not a competition between the scales. Both peaks increase
from $\rho = 0.35$ to $\rho = 0.425$, while the  valley between them decreases. This is indicated by the red arrows.
The green arrows shows that from $\rho = 0.425$ to $\rho = 0.525$ the peaks decrease and the valley increases.
Therefore, the system becomes more structured and then more disordered, which explain the second structural anomaly. 
Related to this transition from disordered-ordered-disordered structured, 
the slope of the MSD curve decreases and then increases, as is shown in (c). The snapshots in (d) show the disks
conformation, including a kagome lattice at $\rho = 0.60$.}
\label{fig4}
\end{figure}

In the anomalous region
the first peak of the RDF increases with
the density while the second peak decreases ~\cite{Barraz09}.
This behavior  is 
also observed  in the 
figure~\ref{fig4}(a) which corresponds 
to the low density and low pressure region of
the figure~\ref{fig3}.  In 
this region as $\rho$ increases, particles move from the 
second at $\approx 2.0$  to the first length scale at 
$\approx 1.2$. Therefore, the system has competition between the scales and, 
as consequence, waterlike anomalies.

The figure~\ref{fig4}(b) shows the RDF
for densities bellow, inside and above the second anomalous region. As 
the density is increased the peaks of the $g(r_{ij})$ related to the
first and to the second length scales increase, reach 
a maximum, and then decrease as the density
is increased. Therefore, even though no competition
between the two length scales is observed, the
system shows density, diffusion and structural anomaly.
In order to reveal the origin of the anomalies, instead
of looking to the two peaks it is necessary to exam
the valley between them.
As the 
density increases from $\rho = 0.35$ to $\rho = 0.425$ this valley goes
down, becoming zero. These zero for the RDF suggests
that as the  density $\rho = 0.425$ is approached from lower densities, 
the system is becoming solid, or well structured. The MSD, illustrated in 
the figure~\ref{fig4}(c), supports this result. The 
 slope of the MSD decreases 
from $\rho = 0.35$ to $\rho = 0.425$ which indicates a decrease in the diffusion. This
is reinforced
by the snapshots shown in the figure~\ref{fig4}(d) as 
an stripe phase. However, increasing
the density of the system even further to $\rho = 0.525$, the disks
becomes disordered. Despite the absence of 
competition, this behavior can also be understood based in the two 
length scales characteristics.

When the fluid is in the stripe structure the interparticle distance
between disks in the same stripe is
the first length scale, and the stripes are separated by the second 
length scale - this is why both peaks increase 
from $\rho = 0.35$ to $\rho = 0.425$. At $\rho = 0.425$ the particles 
have the minimum in the diffusion.
Increasing the density, there is no more space for the stripes remain 
at the distance $\approx 2.0$, and they break 
into the disordered fluid. Essentially, the entalpic contribution to 
the free energy (second length scale) is overcomed
by entropic contribution (the first length scale)~\cite{Bor12}.
Then the system goes from a disordered fluid to a ordered fluid (similar 
to a liquid-crystal) with lower 
diffusion and then gets disordered again, diffusing faster. 
In this reentrant melting region we observe that
$D$ increases with the density while $\uptau$ 
decreases, leading to the second anomalous regions.
As the density increases even more, the system goes to a solid phase 
with a kagome lattice, as the 
snapshot in the figure~\ref{fig4}(d) shows.

Previous studies have shown that the existence of multiples competitive
scales leads to multiples anomalous regions. In the work by Barbosa and 
co-workers~\cite{Barbosa13}
they have used a soft-core potential with three characteristic 
length scales and have
found two TMD lines and transitions between three fluids phases. 
Also, we have observed two structural anomalous regions in quasi-2D systems, 
were the new anomalous regions can be related with the melting of the central layer
between two walls~\cite{BoK15a}. This is similar to what we observed for the 2D system,
were a reentrant melting region leads to the appearance of anomalies.

\section{Conclusion}
\label{conclusions}

Langevin Dynamics simulations of 2D core-softened
disks were performed in order to analyze the system fluid phase
for structural, thermodynamic and dynamic anomalous behavior. 

The core-corona system shows
the presence of two anomalous regions in 
the pressure versus temperature phase diagram. Also, a change in 
the waterlike hierarchy of 
anomalies was observed which can be associated
with the change in the dimensionality.

The two distinct regions with anomalous behavior  observed 
for the colloidal system arises due to two distinct mechanisms.
The first region, at low densities, is 
associated with the competition between the two length scales in the interaction 
potential. This is the same mechanism
observed in previous works and in the molecular system. 
We have shown that the second anomalous region is not related to the competition 
observed in 
the RDF, but to a reentrant fluid phase. This leads the fluid to suffer a 
transition from a disordered structure to a ordered
structure and then back to a disordered structure, resulting in a increase in 
the diffusion as the density increases and a 
decrease of $\uptau$ as $\rho$ increase - the anomalous behavior.

Nevertheless, this was not the first time that we have observed two anomalous 
regions for core-softened fluids. 
In a previous work, core-softened potentials with three scales had lead to two 
regions of density anomaly~\cite{Barbosa13}.
As well, we have showed that fluids modeled by potential equation~(\ref{AlanEq}) 
confined between two flat walls have a
second structural anomaly.
This new anomaly was not related to the competition between the potential 
scales, but
to changes in the number of fluid layers between the walls~\cite{BoK15a}. The 
change in the number of layers
is a additional competition induced by the confinement.
In this work, the competition was induced by the 
resulting fluid reentrant phase.
Therefore, our main finding is that another mechanisms, despite the 
competition between the scales, 
can generate competitions in the system that lead to waterlike anomalies.

\section{Acknowledgments}

We thank the Brazilian agency CNPq for the financial support. JRB 
would like to thanks Professor Alexandre Diehl from Universidade
Federal de Pelotas for the computational time in the TSSC cluster.

%

\end{document}